# D-shaped photonic crystal fiber plasmonic refractive index sensor based on gold grating

Junjie Lu,[1] Yan Li,[1,*] Yanhua Han,[1] Yi Liu,[1] and Jianmin Gao[2]

[1] Department of Optoelectronics Science, Harbin Institute of Technology, Weihai, 264209, China
[2] School of Energy Science and Engineering,, Harbin Institute of Technology, Harbin 150001, China
*Corresponding author: liy@hit.edu.cn

**In this work, we proposed a high resolution D-shaped photonic crystal fiber (PCF) surface plasmon resonance (SPR) sensor based on gold grating. Gold grating is introduced to modulate the resonance wavelength and enhance the refractive index (RI) sensitivity. Structure parameters of PCF and gold grating are analyzed by finite element method (FEM) for optimizing the SPR sensor. The simulation results indicate that air hole pitch, air hole diameter and gold thickness and grating constant have little influence on the sensitivity of refractive index, which reduces the requirement of precise processing. For improving the resolution of RI sensing, two-feature (2F) interrogation method which combines wavelength interrogation and amplitude interrogation is used and the maximum theoretical resolution of SPR sensor reaches to $5.98 \times 10^{-6}$ RIU in range of 1.36-1.38 and the wavelength sensitivity reaches to 3340nm/RIU. The proposed SPR sensor shows potential applications on developing a high-sensitivity, real time and fast-response SPR-RI sensor.**

## 1. INTRODUCTION

Surface plasmon resonance (SPR) is considered as a promising method to detect the tiny refractive index (RI) change, since SPR is extremely sensitive to permittivity of surrounding environment. During the past decade, high sensitivity and real-time detection make the SPR sensor widely used in chemical and biological sensing[1, 2]. Recent years, SPR sensor based on photonic crystal fiber (PCF) has attracted a lot of attention since the first SPR sensor was proposed by Jorgenson in 1993, where the fiber core was coated with the gold film by removing a section of the fiber cladding to exhibit the plasmonic response[3]. PCF bring new vitality to the fabrication of SPR sensors since its unique ability of controlling the evanescent wave penetration. When phase matching condition is met, surface plasmon polaritons (SPP) mode can be excited and results in strong resonance absorption. By detecting the resonance wavelength, unknown analyte RI can be calculated. For investigating the performance of SPR sensor, many PCF structures, such as dual-core PCF[4], D-shaped PCF[5, 6] and birefringent PCF[7], have been well studied in the last decade. For improving the performance of the PCF-SPR sensors, some theoretical designs of metal coated large air hole PCF-SPR sensors have been reported[8, 9]. For instance, Hassani and Skorobogatiy designed a PCF with large air holes, and they coated the inner surfaces of the air holes with metal films to fabricate the PCF-SPR sensor in which the large air holes were used as micro channels. The simulation results showed that the resolution of their sensor could reach $10^{-4}$RIU[10]. For realizing extremely high sensitivity, larger diameter air holes and multichannel analyte-filled PCF structure is proposed. Fan et al. presented a high sensitivity RI sensor based on two large air holes channels filled with analyte which wavelength sensitivity reaches to 7040nm/RIU[11]. Most of the reported PCF SPR sensors are coated with the multiple metal layer and liquid is filled inside the air-holes for increasing the sensitivity, which is difficult in terms of fabrication. Moreover, since the metal film is coated inside the air holes, it is difficult to precisely control the thicknesses of the metal film and it is also time-consuming to fill and re-fill the analyte from the air holes.

In this paper, we proposed a SPR sensor based on D-shaped PCF, which is coated with gold grating for improving the RI sensitivity. Gold grating results in more loss of core mode and by changing the grating constant, the resonance wavelength could be modulated. The resonance wavelength is sensitive to the surrounding refractive index of gold grating, which shows promising potential in RI sensing. Simulation results show that the RI sensitivity is 3 times as much as normal D-shaped PCF-SPR sensor with gold film. By investigating the structure parameters of PCF, simulation results show that PCF structure has little influence on the sensitivity of SPR sensor. Furthermore, the two-feature (2F) interrogation method is used to provide a higher resolution for RI sensing. In RI range of 1.36-1.38, the wavelength sensitivity reaches to 3340nm/RIU, and the maximum theoretical resolution reaches to $5.98 \times 10^{-6}$ RIU, which is much higher than those of wavelength and intensity interrogation method.

## 2. DESIGN AND ANALYSIS

The proposed PCF-SPR sensor is shown in Fig.1 (a). There are two type air holes in the PCF which is side polished with 3μm depth for creating a plane sensing area. Before fabricating the gold grating, gold film should be coated on the plane surface of D-shaped PCF by magnetron sputtering or pulsed laser deposition. Then gold grating can be fabricated by using electron-beam lithography or photolithography. The cross section of the SPR sensor is shown in Fig. 1(b). The diameter of small air holes is $d_s$=0.8μm and the diameter of large air holes is $d_l$=1.6μm, respectively. The lattice pitch between all air holes is Λ=2.3μm. The thickness of the gold gratings is $h_g$=40nm. The gold grating constant is $d_0$=1μm and duty ratio is η=0.5.

The used fiber material is fused silica and the RI is determined by Sellmeier equation [12] as:

$$n(\lambda) = \sqrt{1 + \frac{A_1\lambda^2}{\lambda^2 - B_1} + \frac{A_2\lambda^2}{\lambda^2 - B_2} + \frac{A_3\lambda^2}{\lambda^2 - B_3}}, \quad (1)$$

where $A_1$=0.696166300, $A_2$=0.407942600, $A_3$=0.897479400, $B_1$=4.67914826×$10^{-3}$μm$^2$, $B_2$=1.35120631×$10^{-2}$μm$^2$ and $B_3$=97.9340025μm$^2$. The complex dielectric constant of gold is given by Johnson and Christy[13]. The confinement loss is calculated by using the imaginary part of the effective RI:

$$\alpha = 8.686 \times k_0 \times \text{Im}(n_{eff}) \times 10^4 \text{ dB/cm}, \quad (2)$$

where $k_0 = 2\pi/\lambda$ is the wavenumber and Im($n_{eff}$) is the imaginary part of effective index. The Gaussian beam propagates along the z-direction and mode analysis is performed in the XY plane.

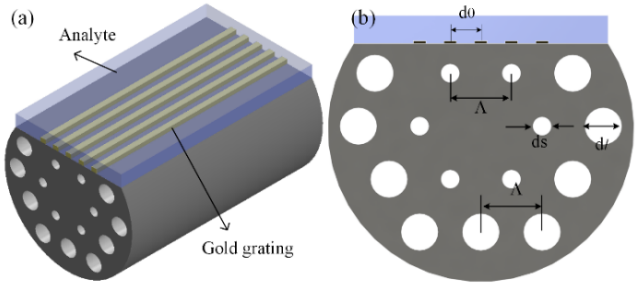

**Fig. 1.** Schematic of proposed PCF-SPR sensor (a) 3D view (b) cross-section of sensor

We investigated the sensitivity difference between SPR sensor with gold film and gold grating. As shown in Fig. 2 (a), when the thickness of gold is 40nm and the RI of analyte changes from 1.36 to 1.37, the resonance wavelength of SPR sensor with full coated gold film shifts from 1389.40nm to 1397.14nm, which means the wavelength sensitivity is 774nm/RIU which is not ideal for RI sensing. The electric field of full coated gold film SPR sensor is shown in Fig. 2(b). Since the wavelength of common commercial broadband laser source is less than 2μm, resonance peak more than 2μm is abandoned. However, when the gold film is replaced with gold grating, which could be fabricated by laser direct writing or photolithography, the resonance wavelength shifts significantly as shown in Fig. 2(a). The wavelength difference is 33.54nm as the resonance wavelength shifting from 1562.56nm to 1596.10nm, which means the wavelength sensitivity is 3354nm/RIU. The simulation result shows that by introducing the gold grating, the wavelength sensitivity of the SPR sensor is 4 times of normal gold film SPR sensor. Moreover, the amplitude sensitivity of gold film SPR sensor is -2.66RIU$^{-1}$ and the gold grating SPR sensor is -14.61RIU$^{-1}$. The gold grating can enhance the loss of the core mode which results in high amplitude sensitivity.

The electric field of gold grating SPR sensor is shown in Fig. 2(c), which resonance wavelength is 1562.56nm. Besides the y-polarized core mode we also can see the SPP mode appears on the interface of gold grating and dielectric as shown in Fig. 2 (d). There is strong coupling between core mode and SPP mode, when the resonance wavelength is 1562.56nm. Therefore, strong coupling results in high confinement loss and high amplitude sensitivity.

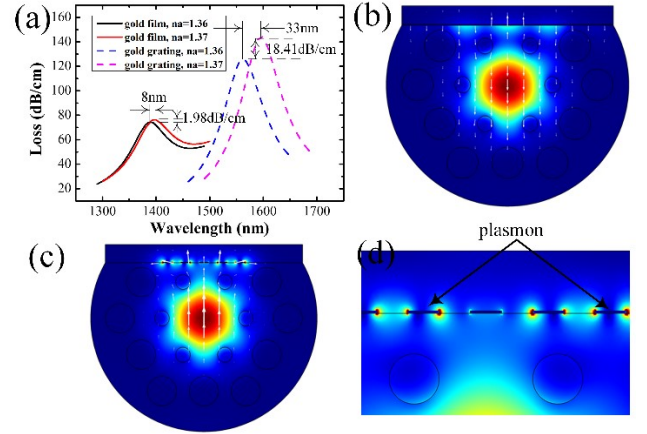

**Fig. 2.** (a) Resonance wavelength of gold film SPR sensor and gold grating SPR sensor with RI from 1.36 to 1.37. (b) The electric field of full coated gold film SPR sensor. (c) The electric field of gold grating SPR sensor. (d) Intensity distribution of SPP mode.

## 3. RESULTS AND DISCUSSIONS

Considering the practical fabrication of the PCF, the diameter of air holes is investigated. The confinement loss spectra is shown in Fig. 3(a), when the diameter of small holes ($d_s$) changes from 0.7μm to 0.9μm and the maximum loss changes from 149.83dB/cm to 173.48 dB/cm. The wavelength sensitivity could be calculated as[14]:

$$S_w = \frac{\Delta\lambda_{peak}}{\Delta n_a}, \quad (3)$$

where $\Delta\lambda_{peak}$ is the difference between two resonance wavelength and $\Delta n_a$ is the variation of analyte RI. As shown in Table 1, different diameter of small air hole has little influence on the wavelength sensitivity. Moreover, we also investigated the sensitivity variation, when the diameter of large air holes ($d_l$) changes. Fig.3 (b) illustrates the loss spectra of gold grating SPR sensor with different diameters of large air holes. With increasing of the diameter, the loss slightly deduces and the resonance wavelength shifts to shorter wavelength. But the diameter of large air hole has limited influence on the wavelength sensitivity as shown in Table 2. The lattice pitch (Λ) also was investigated and the loss spectra is shown in Fig.3 (c). With increasing of the pitch, the resonance wavelength shows slightly blue shift and confinement loss decreases. The reason is that large pitch results in changing of the phase matching condition. As shown in Table 3, different lattice pitch has slightly influence on wavelength sensitivity. The parameters of gold grating also play a vital role in performance of PCF-SPR sensors. As shown in Fig.3 (d), with increasing of the gold grating thickness, more energy of core mode is used for overcoming the damping loss and the resonance wavelength shifts towards to shorter wavelength. When the thickness of gold grating increases, the sensitivity slightly increases as shown in Table 4. To optimize the performance of gold grating SPR sensor, the resonance wavelength is selected around 1550nm and the structure parameters of PCF this should be $d_s$=0.8μm, $d_l$=1.6μm, Λ=2.3μm and the thickness of gold should be $h_g$=40nm.

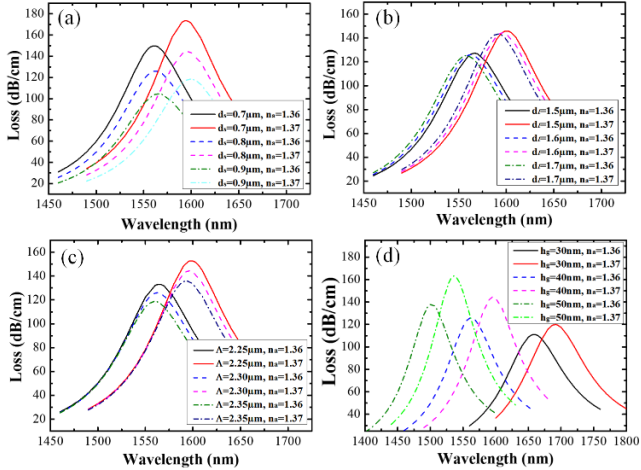

**Fig. 3.** Variation of confinement loss as a function of wavelength with varying (a) diameter of small air holes $d_s$, (b) diameter of large air hole $d_l$ (c) lattice pitch of air holes Λ and (d) gold grating thickness $h_g$, when the RI changes from 1.36 to 1.37

**Table 1.** Relationship between $d_s$ and wavelength sensitivity ($S_w$)

| $d_s$ (μm) | 0.7 | 0.8 | 0.9 |
|---|---|---|---|
| $S_w$ (nm/RIU) | 3320 | 3354 | 3390 |

**Table 2.** Relationship between $d_l$ and wavelength sensitivity ($S_w$)

| $d_l$ (μm) | 1.5 | 1.6 | 1.7 |
|---|---|---|---|
| $S_w$ (nm/RIU) | 3382 | 3354 | 3328 |

**Table 3.** Relationship between Λ and wavelength sensitivity ($S_w$)

| Λ (μm) | 2.25 | 2.3 | 2.35 |
|---|---|---|---|
| $S_w$ (nm/RIU) | 3368 | 3354 | 3344 |

**Table 4.** Relationship between $h_g$ and wavelength sensitivity ($S_w$)

| $h_g$ (nm) | 30 | 40 | 50 |
|---|---|---|---|
| $S_w$ (nm/RIU) | 3244 | 3354 | 3412 |

The gold grating parameters were also investigated for improving the performance of the SPR sensor. As shown in Fig. 4 (a), different grating constant ($d_0$) results in different resonance wavelength, and the wavelength sensitivity also changes as shown in Table 5. We can also observe that with increasing of grating constant, the loss increases which results in higher amplitude sensitivity. Therefore, the grating constant can be used to modulate the resonance wavelength. It is possible for SPR sensors working in expected wavelength such as common commercial broadband laser. Figure 4 (b) depicts the loss spectra of gold grating with different duty ratio (η). We can see that with increasing of duty ratio, the confinement loss increases significantly. When the duty ratio increases, the full width at half maximum (FWHM) reduces, which can get a better signal-to-noise (SNR) ratio. However, the sensitivity slightly decreases as shown in Table 6. Thus, the grating constant is set as $d_0$=1μm and the duty ratio could be set as η=0.5.

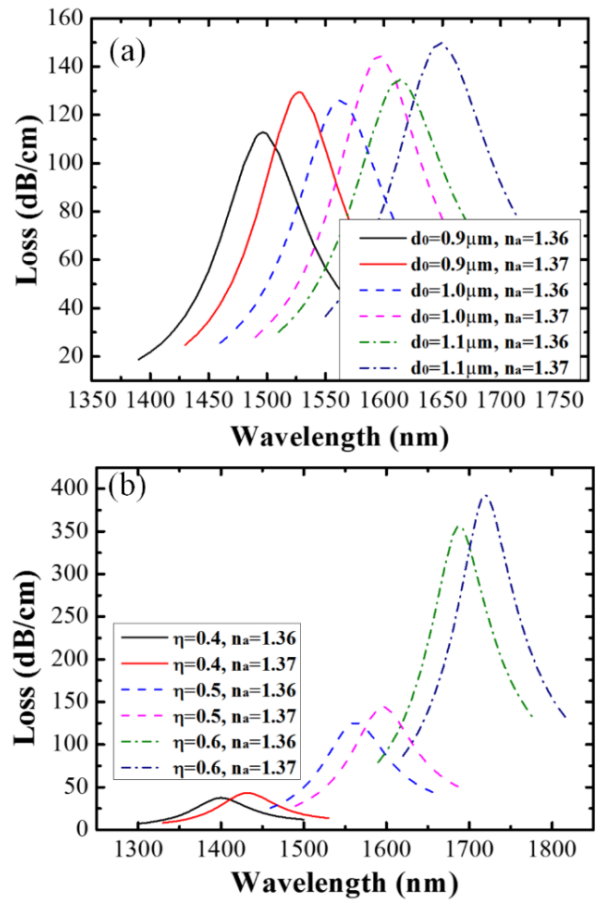

**Fig. 4.** Variation of confinement loss with varying (a) grating constant $d_0$, (b) duty ratio η, when the RI changes from 1.36 to 1.37.

**Table 5.** Relationship between $d_0$ and Wavelength sensitivity ($S_w$)

| $d_0$ (μm) | 0.9 | 1.0 | 1.1 |
|---|---|---|---|
| $S_w$ (nm/RIU) | 3108 | 3354 | 3564 |

**Table 6.** Relationship between η and Wavelength sensitivity ($S_w$)

| η | 0.4 | 0.5 | 0.6 |
|---|---|---|---|
| $S_w$ (nm/RIU) | 3196 | 3354 | 3176 |

Since the SPR sensor is extremely sensitive to the change of surrounding environment RI, we investigated the transmission spectra by changing the analyte RI. From Fig. 5(a) we can see that with increasing of the RI the resonance wavelengths are red-shifted. Therefore, the analyte RI could be detected by measuring the shift of resonance wavelength. As shown in Fig. 5(b), with increasing of analyte RI, the resonance wavelength shows good linear response and the wavelength sensitivity reaches to 3340nm/RIU. In addition, with increasing of analyte RI, the effective index of SPP mode is close to that of core mode, which results in more energy loss of core mode. Therefore, the amplitude also can be used to sensing the RI as a convenient and cost effective method. The amplitude sensitivity can be defined from[15]:

$$s_A \left( \text{RIU}^{-1} \right) = -\frac{1}{\alpha(\lambda, n_a)} \frac{\partial \alpha(\lambda, n_a)}{\partial n_a}, \quad (4)$$

where $\alpha(\lambda, n_a)$ is the loss at RI of $n_a$ and $\partial\alpha(\lambda, n_a)$ is the loss difference between two adjacent analyte RIs.

Figure 5 (c) shows the amplitude sensitivity of the proposed SPR sensor calculated by Eq. (4). From Fig. 5(c), maximum amplitude sensitivity can be obtained at the wavelength of 1660nm, which is -69.3RIU$^{-1}$. Based on the theory mentioned in Ref.[6], the two-feature (2F) sensitivity combines wavelength sensitivity and amplitude sensitivity for enhancing the RI sensitivity. As embedded image shown in Fig.5 (a), if we detect the RI difference Δn of analyte and resonance wavelength shifts from A to B, the 2F sensitivity can be defined as:

$$S_{2F} = \frac{\overline{AB}}{\Delta n} = \frac{\sqrt{\Delta x^2 + \Delta y^2}}{\Delta n}. \quad (5)$$

By normalizing Δx and Δy, we can make wavelength sensitivity and amplitude sensitivity into one dimension. For example, when analyte RI changes from 1.37 to 1.375, wavelength shifts from 1596.10nm to 1612.44nm and loss shifts from 144.42dB/cm to 153.24dB/cm, which means Δx=16.34nm, Δy=8.82dB/cm and Δn=0.005, respectively. We assume the length of sensing area is 1cm. If the wavelength resolution is 0.02nm (AQ3617B) and amplitude resolution is 0.05dB, the RI change will bring 817 data intervals for the x-axis ($DI_x$=16.34/0.02=817) and 176 data intervals for the y-axis ($DI_y$=8.82/0.05=176). Therefore, the 2F data intervals $DI_{2F}$=(817$^2$+176$^2$)$^{1/2}$=836. Finally, the maximum theoretical resolution of wavelength ($R_w$), amplitude ($R_a$) and $R_{2F}$ can be calculated by

$$R = \frac{\Delta n}{DI} = \frac{\Delta n}{\sqrt{DI_x^2 + DI_y^2}}. \quad (6)$$

Results show that 2F resolution ($R_{2F}$=5.98×10$^{-6}$RIU) is higher than theoretical wavelength resolution ($R_w$=6.12×10$^{-6}$ RIU) and theoretical amplitude resolution($R_a$=2.84×10$^{-5}$ RIU).

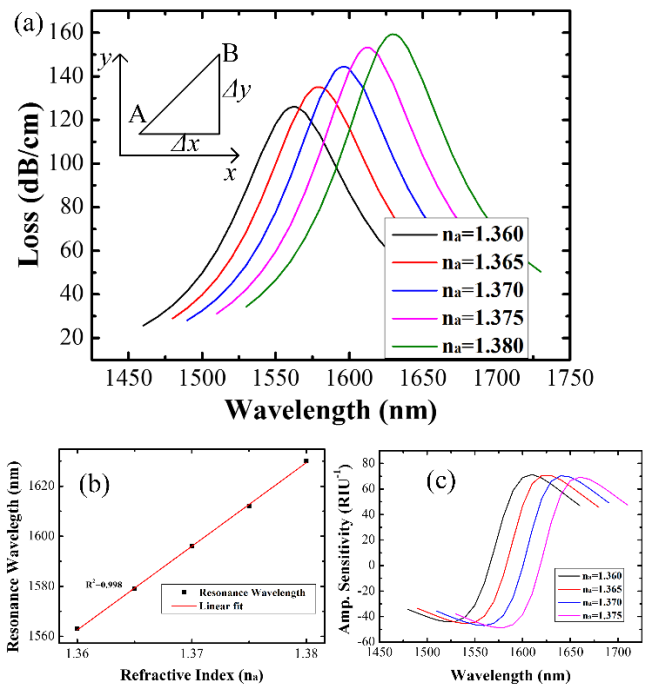

Fig. 5. (a) Variation of confinement loss as a function of wavelength with different analyte RI.(b) Wavelength sensitivity with the analyte RI range from 1.36 to 1.38, (c) amplitude sensitivity with the variation of analyte RI.

## 4. CONCLUSION

A practical D-shaped photonic crystal fiber RI sensor based on surface plasmon resonance is demonstrated by putting gold grating on the flat plane. FEM has been used to analyte the performance of this proposed SPR sensor. The simulation results show that PCF structure parameters have little influence on the performance of sensor. The gold grating is used to enhance the sensitivity and tune the resonance wavelength. Using the 2F interrogation method, the maximum theoretical resolution of RI is improved significantly and reaches to 5.98×10$^{-6}$RIU, which is much higher than wavelength and amplitude resolution. Driving on the advantage of latest nanofabrication technique, the proposed structure can be utilized for environmental, biological and biochemical sensing applications.

**Funding Information.** National Key Research and Development Program of China (Grant No. 2017YFF0209801); Natural Science Foundation of Shandong Province (Grant No. ZR2018MF031); National Natural Science Foundation of China (Grant No. 11504070).